\documentstyle[12pt]{article}
\catcode`\@=11
\def\marginnote#1{}
\newcount\hour
\newcount\minute
\newtoks\amorpm
\hour=\time\divide\hour by60
\minute=\time{\multiply\hour by60 \global\advance\minute by-
\hour}
\edef\standardtime{{\ifnum\hour<12 \global\amorpm={am}%
    \else\global\amorpm={pm}\advance\hour by-12 \fi
    \ifnum\hour=0 \hour=12 \fi
    \number\hour:\ifnum\minute<100\fi\number\minute\the\amorpm}}
\edef\militarytime{\number\hour:\ifnum\minute<100\fi\number\minute}
\def\draftlabel#1{{\@bsphack\if@filesw {\let\thepage\relax
  \xdef\@gtempa{\write\@auxout{\string
    \newlabel{#1}{{\@currentlabel}{\thepage}}}}}\@gtempa
    \if@nobreak \ifvmode\nobreak\fi\fi\fi\@esphack}
     \gdef\@eqnlabel{#1}}
\def\@eqnlabel{}
\def\@vacuum{}
\def\draftmarginnote#1{\marginpar{\raggedright\scriptsize\tt#1}}
\def\draft{\oddsidemargin -.5truein
        \def\@oddfoot{\sl preliminary draft \hfil
        \rm\thepage\hfil\sl\today\quad\militarytime}
        \let\@evenfoot\@oddfoot \overfullrule 3pt
        \let\label=\draftlabel
        \let\marginnote=\draftmarginnote

\def\@eqnnum{(\theequation)\rlap{\kern\marginparsep\tt\@eqnlabel}%
\global\let\@eqnlabel\@vacuum}  }
\def\preprint{\twocolumn\sloppy\flushbottom\parindent 1em
        \leftmargini 2em\leftmarginv .5em\leftmarginvi .5em
        \oddsidemargin -.5in    \evensidemargin -.5in
        \columnsep 15mm \footheight 0pt
        \textwidth 250mmin      \topmargin  -.4in
        \headheight 12pt \topskip .4in
        \textheight 175mm
        \footskip 0pt

\def\@oddhead{\thepage\hfil\addtocounter{page}{1}\thepage}
        \let\@evenhead\@oddhead \def\@oddfoot{} \def\@evenfoot{}
}
\def\titlepage{\@restonecolfalse\if@twocolumn\@restonecoltrue\onecolumn
     \else \newpage \fi \thispagestyle{empty}\c@page\z@
        \def\thefootnote{\fnsymbol{footnote}} }
\def\endtitlepage{\if@restonecol\twocolumn \else  \fi
        \def\thefootnote{\arabic{footnote}}
        \setcounter{footnote}{0}}  
\catcode`@=12
\relax
\def\beq{\begin{equation}}
\def\eeq{\end{equation}}
\def\Im{\mathop{\rm Im}}
\def\NP#1#2#3{{\it Nucl. Phys.} {\bf{#1}} (19#2) #3}
\def\ov{\overline}
\def\PL#1#2#3{{\it Phys. Lett.} {\bf{#1}} (19#2) #3}
\def\PR#1#2#3{{\it Phys. Rev.} {\bf{#1}} (19#2) #3}

\def\Re{\mathop{\rm Re}}
\def\Tr{\mathop{\rm Tr}}

\def\dalpha{{\dot\alpha}}

\def\crbig{\\\noalign{\vspace {3mm}}}
\def\bigint{{\displaystyle\int}}
\def\Fcomp{{\theta\theta}}
\def\Fbarcomp{\ov{\theta\theta}}
\def\Dcomp{{\theta\theta\ov{\theta\theta}}}
\def\Dint{{\bigint d^2\theta d^2\ov\theta\,}}
\def\Fint{{\bigint d^2\theta\,}}
\def\Fbarint{{\bigint d^2\ov\theta\,}}
\def\ex{{\rm exp}}
\relax

\begin{document}
\topmargin-2.4cm
%
\begin{titlepage}
\begin{flushright}
CERN-TH/95-7 \\
NEIP-95-01 \\
IEM-FT-98/95 \\
McGill-95/03 \\
hep--th/9501065 \\
\end{flushright}
\vskip 0.3in
\begin{center}{\Large\bf
Gaugino Condensates and Chiral-Linear Duality:
an Effective Lagrangian Analysis
\footnote{Work supported in part by the Swiss National Foundation,
the European Union (contracts SC1$^*$-CT92-0789 and CHRX-CT92-0004)
CICYT of Spain
(contract AEN94-0928), NSERC of Canada and FCAR of Qu\'ebec.}  }
\vskip 1.3cm
{\bf C. P. Burgess\footnote{
Physics Department, McGill University, 3600 University St., Montr\'eal,
Canada, H3A 2T8},
J.-P. Derendinger\footnote{
Institut de Physique, Universit\'e de Neuch\^atel, A.-L. Breguet 1,
CH--2000 Neuch\^atel, Switzerland}, F. Quevedo$^\ddagger$}
{and}
{\bf M. Quir\'os\footnote{
CERN, TH Division, CH--1211 Gen\`eve 23, Switzerland}
\footnote{On leave from Instituto
de Estructura
de la Materia, CSIC, Serrano 123, E--28006 Madrid, Spain}}
\end{center}
\vskip.5cm
\begin{center}
{\bf Abstract}
\end{center}
\begin{quote}
We show how to formulate the phenomenon of gaugino
condensation in a super-Yang-Mills theory with a field-dependent
gauge coupling described with a linear multiplet. We prove
the duality equivalence of this approach with the more familiar
formulation using a chiral superfield.
In so doing, we resolve a longstanding
puzzle as to how a linear-multiplet formulation can be consistent
with the dynamical breaking of the Peccei-Quinn symmetry which is
thought to occur once the gauginos condense.
In our approach, the
composite gauge degrees of freedom are described by a real vector
superfield, $V$, rather than the chiral superfield that is obtained
in the traditional dual formulation. Our dualization, when applied
to the case of several condensing gauge groups, provides strong evidence
that this duality survives strong-coupling effects in string theory.

\end{quote}
\vskip2.5cm

\begin{flushleft}
CERN-TH/95-7 \\
January 1995 \\
\end{flushleft}

\end{titlepage}
\setcounter{footnote}{0}
\setcounter{page}{0}
\newpage
%
%
\section{Introduction}

Dynamical supersymmetry breaking has been extensively
studied at the non-per\-tur\-ba\-tive level in $N=1$ super-Yang-Mills
theories with and without matter \cite{AKMRV}. It is, in particular,
known that supersymmetry does not break in the theory without
matter. An early analysis of this phenomenon in the pure
super-Yang-Mills theory was given by Veneziano
and Yankielowicz \cite{VY}, whose method has been since extended to
various renormalizable models with charged
matter\footnote{See for instance refs.~\cite{others}
and the review article \cite{AKMRV}.}.

The treatment of (non renormalizable) models having a field-dependent
gauge coupling is less straightforward. These models have the additional
complication that the gauge coupling is induced by expectation
values of scalar fields, which are themselves dynamically generated.
This is notably
the  case for superstring theories, where the gauge coupling is
determined by the expectation value of the dilaton field, whose
potential arises dynamically by gaugino condensation in a hidden
confining gauge sector \cite{DIN,DRSW,condens2,KP}.  The understanding
of gaugino condensation and  supersymmetry breaking in these theories
turns on the description of the dilaton sector.

The supersymmetric partners of the dilaton in the gravity sector of
superstrings are an antisymmetric tensor $b_{\mu\nu}$, with gauge symmetry
$b_{\mu\nu}\longrightarrow b_{\mu\nu}+\partial_{[\mu}b_{\nu]}$
and a Majorana spinor
$\chi$. Since an antisymmetric tensor is by duality equivalent to a
pseudoscalar $\sigma$, there are two dual kinds of descriptions
for the dilaton supermultiplet. Using the original
variable, $b_{\mu\nu}$, leads to a linear superfield $L$ \cite{linear},
while the pseudoscalar, $\sigma$, arises in a chiral superfield, $S$,
in which $\sigma = \Im s$ is the imaginary  part of the lowest scalar
component, $s$. The pseudoscalar that is obtained in this way enjoys
a classical `Peccei-Quinn' (PQ) symmetry of the form
\beq
\label{Sshift}
S \longrightarrow S +i\alpha, \qquad \alpha :
{\rm  a \,\,real\,\, constant},
\eeq
which is dual to the gauge symmetry acting on the antisymmetric tensor.
The transformation from one of these representations to the other is
`chiral-linear' duality.

Past examinations of the problem of gaugino condensation with a
field-dependent gauge coupling have followed ref.~\cite{DRSW} and
have used the chiral-superfield representation of the dilaton.
The formulation of gaugino condensation using directly the linear
multiplet faces an apparent obstacle. The difficulty lies with performing
the chiral-linear duality transformation starting with the chiral
superfield, $S$. This duality transformation requires as its starting
point the existence of the classical symmetry (\ref{Sshift}), but
this symmetry is apparently broken by anomalies in the strongly-coupled
gauge sector. Such considerations have led some workers to entertain the
possibility that the dual theories are {\it inequivalent} once
nonperturbative effects are considered.

In this article we readdress this problem in view of the recent progress
in the understanding of the gauge sector of the effective supergravity
of superstrings due to string loop calculations in $(2,2)$ models
\cite{22,effsugra,DQQ}.  We show how to perform the
condensation analysis directly in terms of the linear multiplet, and
we demonstrate how chiral-linear duality survives gaugino condensation.
We take as
our vehicle for demonstration the cases where a gauge sector with a simple
gauge group condenses, as well as the more general situation where several
commuting factors of the gauge group separately condense.

We start our discussion with the simplest case of a globally
supersymmetric model with a simple gauge group
and without charged matter. We then extend
our results to several gaugino condensates (non-simple gauge groups).
In the string context, these could represent the hidden sector
of a $(2,2)$ model. The introduction of
additional gauge-singlet matter
(moduli) and the generalization
to supergravity are reasonably straightforward \cite{BDQQ}.

It should be emphasized that we do not expect
supersymmetry to break in this simple theory. The effective
potential for gaugino condensates with a field-dependent gauge
coupling described by a chiral multiplet and no further matter
exhibits a `runaway' behaviour, towards the zero-coupling limit.
The fact that chiral-linear duality survives gaugino condensation
implies that the same behaviour will be obtained using the linear
multiplet as starting point.

\section{Duality}

We start with a discussion of duality for the underlying
microscopic theory, well above the condensation scale. The two
representations of the dilaton supermultiplet are a chiral superfield,
$S$, and a linear superfield,  $L$. $S$ satisfies the chiral constraint
$\ov{\cal D}_\dalpha S = 0$ and has as components a complex scalar, $s$,
a spinor $\psi_s$ and a complex auxiliary field, $f_s$.
The real linear superfield \cite{linear}, on the other hand, solves the
constraints
$$
\ov{\cal DD}L = {\cal DD}L = 0,
$$
and has as particle content a real scalar, $C$, an antisymmetric tensor,
$b_{\mu\nu}$ (which only appears through its curl
$
v_\mu = {1\over\sqrt2}\epsilon_{\mu\nu\rho\sigma}\partial^\nu
b^{\rho\sigma}
$),
and a Majorana spinor $\chi$.
The description of a field-dependent gauge coupling with a
linear superfield requires the introduction of the Chern-Simons superfield
$\Omega$, which can be defined by its relation with the chiral superfield of
the Yang-Mills field strengths:
\beq
\label{CSdef}
\Tr(W^\alpha W_\alpha) = \ov{\cal DD} \Omega  ,
\qquad
\Tr(\ov W_\dalpha \ov W^\dalpha) = {\cal DD}
\Omega.
\eeq
Here, $W^\alpha =-{1\over4}\ov{\cal DD} (e^{-{\cal A}}{\cal D}_\alpha
e^{\cal A})$ and
${\cal A}$ is the matrix-valued gauge vector superfield\footnote{
The normalization is $\Tr({\cal A}^2) = \sum_a {\cal A}^a{\cal A}^a$.
}. These conditions define $\Omega$ up to the addition of a real linear
superfield $\Omega_L$.
Contrary to $\Tr(W^\alpha W_\alpha)$, the Chern-Simons superfield is not
gauge invariant. Since its gauge transformation $\delta\Omega$ satisfies
${\cal DD} \delta\Omega=\ov {\cal DD} \delta\Omega = 0$,
$\delta\Omega$ is a real linear superfield and $\Omega_L$
is in some sense a gauge artefact. To construct invariant couplings,
one postulates that the gauge transformation of the linear multiplet is
$$
\delta L  = 2\delta\Omega,
$$
so that the combination
$$
\hat L = L-2\Omega
$$
is gauge invariant. The natural physical dimension of $L$ and $\Omega$
is (energy)$^2$.

A supersymmetric gauge-invariant lagrangian for $L$ takes the form
\beq
\label{Llagr}
{\cal L}_L = 2\mu^2\int d^2\theta d^2\ov\theta \, \Phi(\hat L/\mu^2),
\eeq
where $\Phi$ is an arbitrary real function and $\mu$ is a scale parameter.
If, for instance, ${\cal L}_L$ is the low-energy Wilson effective
lagrangian for a superstring, in the global supersymmetry limit, then the
scale $\mu$ can be regarded as the ultra-violet cutoff which defines the
Wilson lagrangian.

The action of (\ref{Llagr}) contains
kinetic terms for the components of both superfields $L$
and ${\cal A}$. The gauge kinetic terms are
$$
-{1\over2}\Phi_x\,\Tr(F_{\mu\nu} F^{\mu\nu}), \qquad
\Phi_x = \left[ {d\over dx}\Phi(x)\right]_{x= C\mu^{-2}},
$$
which indicates that the gauge coupling constant is
\beq
\label{coupl1}
{1\over g^2} = 2 \Phi_x.
\eeq
It follows that the gauge coupling is a function of the real scalar
field $C$.

We may now dualize this lagrangian to obtain its equivalent in terms
of $S$. To do so, we rewrite (\ref{Llagr}) in an equivalent form by
introducing a real vector superfield $V$, and replacing (\ref{Llagr}) by
\beq
\label{lagr2}
{\cal L}_{V+L} = 2\mu^2\int d^2\theta d^2\ov\theta\,
\Phi\left({V+L\over\mu^2}\right)
+\left({1\over4}\int d^2\theta\, S\ov{\cal DD}(V +2\Omega) +
{\rm h.c.} \right),
\eeq
where $S$ is a chiral superfield. This new theory is
invariant under the gauge transformation
\beq
\label{Vgauge}
V \longrightarrow V+\Delta_L , \qquad
L \longrightarrow L-\Delta_L,
\eeq
where $\Delta_L$ is an arbitrary linear superfield,
$\ov{\cal DD}\Delta_L ={\cal DD}\Delta_L=
0$. To verify the equivalence with (\ref{Llagr}), observe
that the elimination of $S$ imposes the constraint
$$
\ov{\cal DD}(V +2\Omega) = {\cal DD}(V +2\Omega) = 0,
$$
which in turn indicates that $V+2\Omega = V_L$, a linear multiplet.
Gauge invariance (\ref{Vgauge}) allows then the choice
$V_L=0$, which leads again to theory (\ref{Llagr}).
The gauge invariance (\ref{Vgauge}) is a useful tool for the
identification of the
composite degrees of freedom participating in the effective lagrangian.

To dualize, we instead start by removing $V$ and $L$ in favour of
$S+\ov S$. Using
$$
{1\over4}\Fint S\ov{\cal DD} V +{\rm h.c.} =
{1\over4}\Fint S\ov{\cal DD} (V+L)
+{\rm h.c.}= -\Dint (S+\ov S)(V+L),
$$
to rewrite (\ref{lagr2}), we may in principle perform the integration
over $V+L$.  Classically, this involves solving the equation of motion
for the vector superfield $V+L$ in theory (\ref{lagr2}), which is
\beq
\label{VLeom}
2\left[ {d\over dX} \Phi(X) \right]_{X={V+L\over\mu^2}} = S+\ov S.
\eeq
$V+L$ is thereby given implicitly as a function of $S+\ov S$:
$X(S+\ov S) \equiv (V+L)/\mu^2$. The dual theory is then
\beq
\label{Ldual}
{\cal L}_{dual} =
\mu^2\Dint K(S+\ov S) +{1\over2}\left( \Fint S\Tr(W^\alpha W_\alpha) +
{\rm h.c.} \right),
\eeq
with $K(S+ \ov S)$ the result of integrating over $V+L$. Classically:
\beq
\label{Kis}
K(S+\ov S) = \left[ 2\Phi(X) - (S+\ov S) X \right]_{X=X(S+\ov S)} .
\eeq
Notice that the lowest component of
(\ref{VLeom}) gives the equality of the two dual expressions for
the field-dependent gauge coupling:
\beq
\label{couplS}
{1\over g^2} = 2\Re s = 2\Phi_x.
\eeq

\section{Effective Actions}

We now turn to a study of gaugino condensation in theory (\ref{Llagr}).
To proceed, we firstly consider the more familiar case of the
chiral theory. With chiral multiplets only, a field-dependent
gauge coupling can always be introduced with the lagrangian
\beq
\label{SYM1}
{\cal L} = {\cal L}_S +{1\over2}\Fint S\Tr(W^\alpha W_\alpha)
+{1\over2}\Fbarint \ov S\Tr(\ov W_\dalpha \ov W^\dalpha),
\eeq
The kinetic terms of $S$ are contained in ${\cal L}_S$ as well as possible
contributions from other chiral multiplets (moduli).
Since by assumption the theory does not contain charged matter,
${\cal L}_S$ will not depend on ${\cal A}$.

For a non-abelian gauge group, asymptotic freedom of theory (\ref{SYM1})
leads to a confining regime at a scale where condensates appear.
To study the formation of condensates of gaugino bilinears, we follow
ref.~\cite{CJT} and compute the generating functional $\Gamma$
of two-particle-irreducible (2PI) Green's
functions\footnote
{This is similar, but not identical \cite{BDQQ}, to the approach
of ref.~\cite{VY}.}. In the present instance, this is accomplished
by coupling an external
chiral superfield of currents, $J$, to $\Tr(W^\alpha W_\alpha)$,
which includes gaugino bilinears in its lowest component, as follows:
$$
\ex\left\{i\hat W[J,S]\right\} =
\int {\cal D A}\, \ex\left\{i\int d^4x\Fint \left({1\over2}
S+J \right) \Tr(W^\alpha W_\alpha) +{\rm h.c.} \right\}.
$$
We then Legendre transform from the variable $J$ to the variable $U$,
resulting in the 2PI effective action
$$
\Gamma[U,S] = \hat W[J,S] - \int d^4x\, \left[\Fint UJ +{\rm h.c.}
\right],
$$
where the chiral superfield $U$ is given by
\beq
\label{Udef}
U = {\delta\hat W\over\delta J} = \langle\Tr(W^\alpha W_\alpha)\rangle.
\eeq
Notice that $U$ is not an arbitrary chiral superfield. It follows from
relations (\ref{CSdef}) that a real superfield $\tilde V$ exists such that
$U=-{1\over2}\ov{\cal DD}\tilde V$. Clearly $\tilde V$ is defined up to
the addition of a linear superfield.
Integrating over the gauge superfield ${\cal A}$ leads to the effective
lagrangian for the composite field $U$ and the chiral
field-dependent gauge coupling $S$. One obtains \cite{BDQQ}:
\beq
\label{result}
\begin{array}{rcl}
\Gamma[U,S] &=& \bigint d^4x\, {\cal L}_U , \crbig
{\cal L}_U &=& \Dint K_U(U, \ov U)
+ \Fint w_U(U,S) + \Fbarint \ov w_U(\ov U,\ov S),
\end{array}
\eeq
with a superpotential given by
\beq
\label{superpot1}
w_U(U,S) = {1\over2}SU + {A\over12} U\log \left({ U\over
M^3}\right)
= {A\over12}U\log\left({Ue^{6S/A}\over M^3}\right), \qquad
A={3C(G)\over8\pi^2} .
\eeq
In these expressions, $A$ is the coefficient of the one-loop
gauge beta-function, $C(G)$ the quadratic Casimir of the gauge group
$G$ and $M$ is a cutoff scale.

The result (\ref{superpot1}) was first obtained by Veneziano
and Yankielowicz \cite{VY}, without a field-dependent gauge
coupling ({\it i.e.} $S=$ constant), and by
Taylor \cite{T}, with the chiral superfield $S$. Although nothing which
follows depends on the form taken for the K\"ahler potential $K_U$,
we assume here for concreteness the expression obtained in ref.~\cite{VY}:
$K_U = h(U\ov U)^{1/3}$, with $h$ a dimensionless constant.

It is important to recognize \cite{BDQQ} that $U$ here
is a purely classical field that was obtained by Legendre
transforming the externally-applied current, $J$.
It is determined in terms of $S$ by extremizing $\Gamma$ with
respect to variations of $U$. Once this is done, we can add
the result ${\cal L}_U(S,U(S))$ to ${\cal L}_S$ --- the
${\cal A}$-independent part of the original lagrangian
(\ref{SYM1}) --- and so obtain an effective lagrangian
\beq
\label{Leff}
{\cal L}_{chiral} = {\cal L}_S + {\cal L}_U,
\eeq
which governs the low-energy dynamics of the integration over $S$.

We now repeat this process using the formulation of the dilaton
with a linear multiplet.
As point of departure we take expression (\ref{lagr2}),
but this time using the gauge $L=0$ [or $\Delta_L =L$ in (\ref{Vgauge})].
With relations (\ref{CSdef}), lagrangian (\ref{lagr2}) becomes
\beq
\label{lagr3}
{\cal L}_V = 2\mu^2\bigint d^2\theta d^2\ov\theta\, \Phi(V/\mu^2)
+\left( {1\over2}\bigint d^2\theta\, S \left\{\Tr(W^\alpha
W_\alpha)
+{1\over2}\ov{\cal DD}V \right\} + {\rm h.c.} \right).
\eeq
What is interesting in this last form is that the dependence on the gauge
superfield ${\cal A}$ is entirely in the term linear in $S$. And this term is
identical to gauge kinetic terms in the usual formulation (\ref{SYM1})
of the super-Yang-Mills theory, with a field-dependent
gauge coupling specified by a chiral superfield. The calculation leading to
the effective lagrangian is then identical to the chiral case described above.
Borrowing the result, one obtains
\beq
\label{Lefflin}
\begin{array}{rcl}
{\cal L}_{linear} &=& \Dint \left\{ 2\mu^2 \Phi(V/\mu^2)
+ h(U\ov U)^{1/3} \right\} \crbig
&& +\left( \Fint \left\{{1\over2} S(U+{1\over2}\ov{\cal DD}V)
+{A\over12} U\log \left({ U/M^{3}}\right)\right\} +{\rm h.c.}\right).
\end{array}
\eeq
Recall that the superfield $S$ has been introduced
as a Lagrange multiplier in the original theory (\ref{lagr3}).
It plays the same r\^ole in the effective
lagrangian: integrating over $S$ leads to
\beq
\label{Uis}
U = -{1\over2}\ov{\cal DD}V,
\eeq
and the final form of the effective lagrangian for the linear multiplet
theory (\ref{Llagr}) is
\beq
\label{efflagrfinal}
{\cal L}_{linear} = \Dint \left\{ 2\mu^2 \Phi(V/\mu^2)
+ h(U\ov U)^{1/3} \right\}
+\left\{ \Fint {A\over12} U\log \left(U/M^3\right)
+{\rm h.c.} \right\} ,
\eeq
with $U$ as in eq.~(\ref{Uis}). We have then obtained an
effective theory where
the degrees of freedom of the short distance theory, described by a linear
multiplet with components $(C,\chi,b_{\mu\nu})$ and a gauge superfield
$(\lambda^a, a_\mu^a, D^a)$ (in Wess-Zumino gauge) have been replaced by
the components of a complete real vector superfield $V$ (eight
bosons and eight fermions).

Having now obtained explicit expressions for the effective (2PI)
actions for the low-energy limit of the dual formulations of the
underlying microscopic theory, it is instructive to display the
equivalence under duality at the
level of the effective theories,  ${\cal L}_{ linear}$  of
eq.~(\ref{efflagrfinal})
and ${\cal L}_{ chiral}$ of eqs.~(\ref{result}), (\ref{superpot1}) and
(\ref{Leff}), which combine to give:
\beq
\label{Leffdual}
\begin{array}{rcl}
{\cal L}_{chiral} &=& \Dint[\mu^2 K(S+\ov S)
+h(U\ov{U})^{1/3}] \crbig
&& +\left( \Fint [{1\over2}S U+{A\over12} U\log( U/M^3)]
+{\rm h.c.}\right).
\end{array}
\eeq
To obtain ${\cal L}_{linear}$ from ${\cal L}_{chiral}$, one uses the fact
that the composite superfield $U$ can be obtained from a real vector
superfield $\tilde V$ by $U=-{1\over 2} \ov{\cal DD} \tilde V$.
Then write
\beq
\begin{array}{l}
\mu^2\Dint K + {1\over2}\left(\Fint S U +{\rm h.c.}
\right) =
\Dint \left[\mu^2 K + (S+\ov S)\tilde V \right] \crbig
\hspace{4.5cm}
= \Dint \left[ 2\mu^2\Phi(X) +(S+\ov S)(\tilde V-\mu^2 X) \right] .
\end{array}
\eeq
The last equality uses the microscopic duality result to express
$K$ as a function of $X$. Integration over
$S$ then imposes the constraint $X = \tilde V\mu^{-2}$, leading to
\beq
\label{interm}
\mu^2\Dint K + {1\over2}\left(\Fint S U +{\rm h.c.}
\right) = 2\mu^2 \Dint \Phi(\tilde V/\mu^2),
\eeq
which completes the proof of the equivalence of theories
(\ref{efflagrfinal}) and (\ref{Leffdual}).

We remark that the shift symmetry (\ref{Sshift})
has a macroscopic counterpart in the effective lagrangian (\ref{Leffdual})
because of the constraint (\ref{Uis}) which is satisfied
by the composite chiral superfield $U$. Imposing that transformation
(\ref{Sshift}) generates an anomaly leads to the condition
\begin{equation}
\label{anomalia}
{i\over2}\alpha\left(\Fint U - \Fbarint \ov{U}\right)
= 2\alpha\Im  f_u =
{\rm a \,\, total \,\, derivative\,\,}(\partial^\mu \tilde v_\mu)
\end{equation}
($\alpha$ is a real constant), which in turn implies $U=-{1\over2}
\ov{\cal DD}\tilde V$. The  right-hand-side of eq.~(\ref{anomalia}) is the
macroscopic version of the anomaly. This indicates that symmetry
(\ref{Sshift}) is broken by non-perturbative effects.

\section{Component Expressions}

The physical content of these manipulations becomes clearer once the above
expressions are expanded in terms of the components of the various superfields,
as we do in this section.

The effective lagrangian (\ref{Lefflin}) for the linear multiplet theory
describes eight bosonic and eight
fermionic degrees of freedom. The vector superfield $V$ replaces the
combination $L-2\Omega$, which appears in the microscopic theory (\ref{Llagr}).
It is to be functionally integrated in the
low-energy theory, subject to the constraint, $U=-{1\over2}\ov{\cal
DD}V$, which fixes four bosonic and four  fermionic components in terms of
the components of the chiral field $U = \langle \Tr(W^\alpha W_\alpha)
\rangle$. Those degrees of freedom in
$V$ which are not included in $U$ represent the gauge invariant
completion of the degrees of freedom of
the original linear superfield $L$.  Indeed, the invariance (\ref{Vgauge})
applied to the lagrangian (\ref{lagr2}) indicates that $L$ can be
absorbed in $V$. To study the component expansion of the
effective theory, it is therefore convenient to use an expansion
of $V$ which makes this gauge transformation (\ref{Vgauge})
transparent. If the expansion of the superfield  parameter is
\beq
\label{linexp}
\Delta_L = c_L +i\theta\varphi_L
-i\ov{\theta\varphi}_L
-\theta\sigma^\mu\ov\theta (v_L)_\mu
+{1\over2}\Fcomp\ov\theta \partial_\mu\varphi_L
\sigma^\mu
+{1\over2}\Fbarcomp\theta \sigma^\mu\partial_\mu\ov\varphi_L
+ {1\over4}\Dcomp\Box c_L,
\eeq
with
$$
(v_L)_\mu ={1\over\sqrt2} \epsilon_{\mu\nu\rho\sigma}
\partial^\nu b^{\rho\sigma}_L,  \qquad
[{\rm or}\,\, \partial^\mu (v_L)_\mu = 0],
$$
it is natural to use for $V$ the expansion
\beq
\label{Vexp}
\begin{array}{rcl}
V&=& c +i\theta\varphi -i\ov{\theta\varphi} -{1\over2}\Fcomp \ov m
-{1\over2}\Fbarcomp m
-\theta\sigma^\mu\ov\theta v_\mu \crbig
&&
+{1\over2}\Fcomp\ov\theta(\ov\Lambda+\partial_\mu\varphi\sigma^\mu)
+{1\over2}\Fbarcomp\theta(\Lambda+\sigma^\mu\partial_\mu\ov
\varphi) +\Dcomp(d+{1\over4}\Box c).
\end{array}
\eeq
With this choice, the tranformation (\ref{Vgauge}) shifts $c$ and $\varphi$,
while leaving $m$, $\Lambda$ and $d$ invariant,
Its action on $v_\mu$ is
$$
v_\mu\longrightarrow v_\mu + {1\over\sqrt2}
\epsilon_{\mu\nu\rho\sigma}
\partial^\nu b^{\rho\sigma}_L,
$$
which is the gauge transformation of the three-index antisymmetric tensor
\beq
\label{3index}
v_\mu = {1\over\sqrt2}
\epsilon_{\mu\nu\rho\sigma}h^{\nu\rho\sigma}.
\eeq
The real superfield $V$ with gauge invariance (\ref{Vgauge}) is then the
supersymmetric description of the three-index tensor \cite{G},
while the invariant chiral superfield
$U=-{1\over2}\ov{\cal DD} V$ is the supersymmetrization of its
gauge-invariant curl. Using the expansion
$$
U = u - i\theta\sigma^\mu\ov\theta\partial_\mu u
-{1\over4}\Dcomp \Box u
+\sqrt2\theta\psi_u +{i\over\sqrt2}\Fcomp (\partial_\mu\psi_u
\sigma^\mu\ov\theta) -\Fcomp f_u ,
$$
its components are given by
\beq
\label{UVcomp}
u = -m \,;\,\qquad
\sqrt2\psi_u = \Lambda \,;\, \qquad
f_u = -2d+i\partial^\mu v_\mu =
-2d+{i\over\sqrt2}\epsilon_{\mu\nu\rho\sigma}H^{\mu\nu\rho\sigma},
\eeq
where the curl
$$
\epsilon_{\mu\nu\rho\sigma}
H^{\mu\nu\rho\sigma} = \epsilon_{\mu\nu\rho\sigma}
\partial^{[\mu}h^{\nu\rho\sigma]}
=\sqrt2 \partial^\mu v_\mu
$$
describes a single degree of freedom. The last equation (\ref{UVcomp})
contains an important piece of information: writing a chiral
superfield in the form $U=-{1\over2}\ov{\cal DD}V$ implies that
$\Im f_u ={1\over\sqrt2}\epsilon_{\mu\nu\rho\sigma}
H^{\mu\nu\rho\sigma}$. For all other components, (\ref{Uis}) is a simple
change of variables, from $(m,\Lambda,d)$ to $(u,\psi_u,\Re f_u)$.
In other words, if a chiral superfield $U$ is such that the
imaginary part of its $f_u$ component is a total derivative
$\partial^\mu v_\mu$, then there exists
a vector superfield $V$, defined up to the addition of a linear
superfield, such that $U=-{1\over2}\ov{\cal DD}V$.

Since the fields $c$, $\varphi$ and the transverse
part $v_\mu^\bot$ of the vector $v_\mu$ form the components of a linear
multiplet [see eq.~(\ref{linexp})], they do not appear in (\ref{UVcomp}).
The components of $V$ which appear in eq.~(\ref{UVcomp}) can be expressed in
terms of the underlying composite degrees of freedom, following
(\ref{Udef}). That is,
$U = \langle \Tr(W^\alpha W_\alpha) \rangle$ implies
\beq
\label{Uident}
\begin{array}{rcl}
m & = & \langle \Tr (\lambda\lambda)\rangle, \crbig
\Lambda & = & \langle -2i\Tr(\lambda D)
+ \Tr (F_{\mu\nu}\sigma^\mu \ov\sigma^\nu\lambda)\rangle , \crbig
d & = & \langle\Tr( -{1\over4}F^{\mu\nu}F_{\mu\nu} +
{i\over2}\lambda\sigma^\mu D_\mu\ov\lambda
-{i\over2}D_\mu\lambda\sigma^\mu\ov\lambda +{1\over2} D^2 )
\rangle , \crbig
\partial^\mu v_\mu & = & \langle
\Tr({1\over4} \epsilon_{\mu\nu\rho\sigma} F^{\mu\nu}
F^{\rho\sigma} -\partial_\mu[\lambda\sigma^\mu\ov\lambda])\rangle.
\end{array}
\eeq
We see that the gaugino condensate is described either by the
complex scalar $m$ --- which is the $\ov{\theta\theta}$ component of $V$
--- or the lowest component $u$ of $U$. Notice also that the quantity
$\partial^\mu v_\mu$
(which is constrained to equal the imaginary part of $f_u$) corresponds
as it should to the anomalous divergence of the supersymmetry
chiral current \cite{GW}.

The (gauge-invariant) low-energy states $c$, $\varphi$ and $v_\mu^\bot$,
on the other hand,  correspond to the physical degrees of freedom $C$, $\chi$
and
$b_{\mu\nu}$ of  the linear multiplet present at the microscopic level,
supplemented by the gauge-variant part of the Chern-Simons superfield.
For instance, computing the Chern-Simons superfield in the Wess-Zumino
gauge, the correspondence would be
\beq
\label{ident3}
\begin{array}{rcl}
c &\longleftrightarrow& C,  \crbig
\varphi &\longleftrightarrow& \chi+{i\over2}
\Tr(\sigma^\mu\ov\lambda a_\mu), \crbig
v_\mu &\longleftrightarrow& {1\over\sqrt2}\epsilon_{\mu\nu\rho\sigma}
(\partial^\nu b^{\rho\sigma}+\sqrt2 \omega^{\nu\rho\sigma})
-\Tr(\lambda\sigma^\mu\ov\lambda) ,
\end{array}
\eeq
where the normalization of the bosonic Chern-Simons form is
$\epsilon_{\mu\nu\rho\sigma}\Tr(F^{\mu\nu}F^{\rho\sigma})=4
\epsilon_{\mu\nu\rho\sigma}\partial^\mu\omega^{\nu\rho\sigma}$.
Notice that the expression for the longitudinal part
of $v_\mu$ is compatible with the last equation
(\ref{Uident}). In contrast with the components of $U$, the fields
$c$, $\varphi$ and $v_\mu^\bot$ should be regarded as the quantum fields
of the effective theory far below the condensation scale.

The component expansion of the effective lagrangian (\ref{efflagrfinal}) is
now easily obtained, using eqs.~(\ref{Vexp}) and (\ref{UVcomp}).
It is as usual
the sum of bosonic contributions and terms quadratic or quartic in fermions:
$$
{\cal L}_{linear} =
{\cal L}_{bos.} + {\cal L}_{quad.} + {\cal L}_{quart.}\,.
$$
Using the notation
$$
\Phi_x = \left[{d\Phi(x)\over dx}\right]_{x= c\mu^{-2}}, \qquad
\Phi_{xx} = \left[{d^2\Phi(x)\over dx^2}\right]_{x= c\mu^{-2}}, \qquad\ldots,
$$
the bosonic contributions which determine the vacuum structure
of the theory are:
\beq
\label{bos}
\begin{array}{rcl}
{\cal L}_{bos.} &=&
{1\over2}\mu^{-2}\Phi_{xx}\left[ m\ov m + v^\mu v_\mu - (\partial_\mu
c)(\partial^\mu c)\right] + 2\Phi_x d  \crbig
&& +{h\over9} (m\ov m)^{-2/3} \left[ (\partial_\mu \ov m)(\partial^\mu m)
+4d^2+ (\partial_\mu v^\mu)^2 \right]  \crbig
&& +{A\over6}d \left[ 2+\log({m\ov m\over M^6}) \right] +
{A\over12}i(\partial^\mu v_\mu) \log ({\ov m\over m}).
\end{array}
\eeq
Since a gaugino condensate
$\langle\Tr(\lambda\lambda)\rangle$ is described by
the expectation value of $m$, its phase only appears
in the coupling ${A\over12}i(\partial^\mu v_\mu) \log(\ov m/m)$ which
arises in the effective theory as a consequence of the anomaly of
R-symmetry [see refs.~\cite{VY,GW} and the last equation (\ref{Uident})].
The terms quadratic in fermion fields are:
\beq
\label{quad}
\begin{array}{rcl}
{\cal L}_{quad.} &=&
-{i\over2} \mu^{-2}\Phi_{xx} (\ov\varphi\gamma^\mu\partial_\mu\varphi)
+i{h\over36} (m\ov m)^{-2/3} (\ov\Lambda\gamma^\mu\partial_\mu\Lambda)
\crbig
&& -{1\over4} \mu^{-4}\Phi_{xxx}\left[ m\ov\varphi_R\varphi_L
+ \ov m\ov\varphi_L\varphi_R - v^\mu(\ov\varphi\gamma_\mu\gamma_5\varphi)
\right]
\crbig
&& -{i\over2}\mu^{-2}\Phi_{xx} \left[\ov\varphi_R\Lambda_L -
\ov\varphi_L\Lambda_R \right]
+i{h\over108}(m\ov m)^{-2/3}(\ov\Lambda\gamma^\mu\gamma_5\Lambda)
\partial_\mu(\log {\ov m\over m})
\crbig
&& -{h\over27}(m\ov m)^{-2/3} d\left[ m^{-1}\ov\Lambda_R\Lambda_L
+\ov m^{-1}\ov\Lambda_L\Lambda_R \right]
\crbig
&& -i{h\over54} (m\ov m)^{-2/3}(\partial^\mu v_\mu)
\left[ m^{-1}\ov\Lambda_R\Lambda_L
- \ov m^{-1}\ov\Lambda_L\Lambda_R \right]
\crbig
&& +{A\over48} m^{-1}\ov\Lambda_R\Lambda_L +{A\over48}\ov m^{-1}
\ov\Lambda_L\Lambda_R.
\end{array}
\eeq
Finally, the quartic fermionic terms are simply
\beq
\label{quart}
{\cal L}_{quart.} =
{1\over8}\mu^{-6}\Phi_{xxxx}(\ov\varphi_R\varphi_L)(\ov\varphi_L\varphi_R)
-{19h\over 324} (m\ov m)^{-5/3} (\ov\Lambda_R\Lambda_L)(\ov\Lambda_L\Lambda_R).
\eeq

We may now use this lagrangian to determine the theory's vacuum structure.
The first step is to eliminate the constrained field, $d$, using its equation
of motion. Since $d$ is auxiliary its equation may be solved algebraically,
with solution
\beq
\label{deom}
d = -{9\over8h}(m\ov m)^{2/3} \left\{2\Phi_x+{A\over6}
\left[ 2+\log\left( {m\ov m\over M^6}\right) \right] \right\}
+{1\over24} \left\{ m^{-1}\ov\Lambda_R\Lambda_L +\ov
m^{-1}\ov\Lambda_L\Lambda_R\right\}.
\eeq
Using this to eliminate $d$ in the lagrangian gives the following scalar
potential for the fields $c$ and $m$\footnote{The potential depends
on $c$ via $x=c\mu^{-2}$  and $\Phi=\Phi(x)$.}
\beq
\label{V1}
\begin{array}{rcl}
V_{linear} &=& \displaystyle{
{4h\over9}(m\ov m)^{-2/3}  d^2 -{1\over2}\mu^{-2}\Phi_{xx}m\ov m}
\crbig
&=& \displaystyle{
{9\over16h}(m\ov m)^{2/3}\left\{ 2\Phi_x+{A\over6}\left[2+
\log\left({m\ov m\over M^6}\right)\right]\right\}^2
-{1\over2}\mu^{-2}\Phi_{xx} m\ov m}.
\end{array}
\eeq

Since the quantities $h(m\ov m)^{-2/3}$ and $-\Phi_{xx}$ appear in the
kinetic terms for $m$ and $c$ respectively [see (\ref{bos})], these
quantities must be positive. The scalar potential is therefore the
sum of two non-negative terms, and so is minimized when they both
vanish. We are therefore led to the conditions $d = m = 0$,
leaving only the expectation value of the
lowest component, $c$, undetermined (so far).
The condition $d =0$ implies
\beq
\label{gaugino1}
 m\ov m = M^6 e^{-2}\langle e^{-32\pi^2\phi_x/C(G)}\rangle.
\eeq
Since the gauge coupling constant in the microscopic theory is $g^{-2} =
2\langle\Phi_x\rangle$, which is a function of the free quantity
$\langle c\rangle/\mu^{-2}$, we see that (\ref{gaugino1}) has the usual
form
$$
 |m| = M^3 e^{-1} e^{-8\pi^2 g^{-2}/C(G)}.
$$
Equation (\ref{gaugino1}) exhibits the familiar `runaway' behaviour
of the theory with a field-dependent gauge coupling and no matter.
That is, the minimization
condition $m=0$ now implies that $\langle c \rangle $ prefers to
take values for which
the gauge coupling vanishes, for which condensates do not form and
supersymmetry
does not break.

It is natural to identify the scale $\mu$ which appears in the microscopic
lagangian (\ref{Llagr}) with the ultra-violet cutoff, $M$,
of the macroscopic theory. In this case, the Wilson gauge coupling,
$2\Phi_x$, is the Wilson gauge coupling defined
at scale $M$, and condensates should form at the renormalization-group
invariant scale $M_{cond.}\sim M \,{\rm exp}[-1/Ag^2(M)]$.

To further sharpen the previous discussion concerning the
duality between the two descriptions of the gaugino condensation,
we now compare the above
discussion with the equivalent analysis using the chiral superfield $S$.
For the chiral theory defined by eq.~(\ref{Leffdual}),
the scalar potential (after
eliminating the auxiliary fields $f_u$ and $f_s$) is
\beq
\label{Vdual}
\begin{array}{rcl}
V_{chiral}&=&{\displaystyle {h\over9}(u\ov u)^{-2/3}f_u\ov f_u+
\mu^2 K_{s\ov s}f_s\ov f_s}
\crbig
&=&{\displaystyle {9\over 4h}(u\ov u)^{2/3}\left| s +
{A\over6}[1+\log({u\over M^3})]
\right|^2 + {1\over4}\mu^{-2} K_{s\ov s}^{-1} (u\ov u)},
\end{array}
\eeq
where $s$ and $u$ are the lowest complex scalar components
of the chiral superfields $S$ and $U$. Since $K_{s\ov s} =
{\partial^2\over\partial s \partial\ov s} K(s+\ov s) $
is the kinetic metric for $s$, this potential is once
more the sum of two non-negative terms, and so
is minimized by $f_u = f_s = 0$.

These conditions can be compared to those obtained using the linear multiplet
by using the duality relations (\ref{VLeom}) and (\ref{Kis}),
which imply
$$
K_{s\ov s} = - {1\over2}\left[\Phi_{xx} \right]_{x= x(s+\ov s)}^{-1}.
$$
Since $u\ov u = m\ov m$, the second term in $V_{chiral}$ is
clearly identical to the second term in $V_{linear}$. In contrast with
the similar term in eq.~(\ref{V1}), the first contribution in
$V_{chiral}$ is proportional to the square of the absolute
value of a complex quantity. Cancelling the first term in
$V_{chiral}$ requires
\beq
\label{uis}
u  = M^3 e^{-1}\langle e^{-16\pi^2 s/C(G)}\rangle,
\eeq
an equation relating the {\it complex} quantities $u$ and $s$.
Its absolute value is identical to (\ref{gaugino1}) since
$|u| =|m|$ and $\Re s = \Phi_x$ according to eq.~(\ref{VLeom}).
But the argument of eq.~(\ref{uis}),
\beq
\label{arg}
\langle s-\ov s\rangle -{A\over6}\log (\ov m/m)  = 0,
\eeq
which relates the phase of the gaugino condensate
$m$ to $\langle \Im s\rangle$, does not exist in the
case of the linear multiplet.

To make the comparison explicit, separate $\Re f_u$ and $\Im f_u$
in $V_{chiral}$, and write
\beq
\label{difference}
\begin{array}{rcl}
V_{chiral} &=& {\displaystyle
{9\over16h}(m\ov m)^{2/3} \left\{ s+\ov s+{A\over6} [2+
\log({m\ov m\over M^6}) ] \right\}^2 - {1\over2}\mu^{-2}
\Phi_{xx}m\ov m}
\crbig
&& {\displaystyle
-{9\over16h}(m\ov m)^{2/3}\left\{s-\ov s -{A\over6}
\log({\ov m\over m}) \right\}^2}.
\end{array}
\eeq
Since eq.~(\ref{VLeom}) allows us to rewrite the potential (\ref{V1}) as
\beq
\label{V2}
V_{linear} = {9\over16h}(m\ov m)^{2/3} \left\{ s+\ov s+{A\over6} [2+
\log({m\ov m\over M^6}) ] \right\}^2 - {1\over2}\mu^{-2}\Phi_{xx}m\ov m,
\eeq
one obtains
\beq
\label{difference2}
V_{chiral} = V_{linear} + {h\over9}(m\ov m)^{-2/3}(\Im f_u)^2.
\eeq
The two potentials (\ref{difference}) and (\ref{V2})
differ by the last term in eq.~(\ref{difference}), which leads to
the supplementary vacuum equation (\ref{arg}). Since we have explicitly
proven the equivalence of the two dual theories, see (\ref{interm}),
the information contained in both theories must however
be the same.

To understand the meaning of the vacuum equation
(\ref{arg}), which is equivalent to $\Im f_u =0$,
it is useful to recall two
facts about the chiral effective lagrangian (\ref{Leffdual}). Firstly,
the superpotential (\ref{superpot1}) is invariant under an
R-symmetry rotation of the composite superfield $U$, which rotates the
gaugino condensate by a phase $\beta$, combined with the
shift $S\longrightarrow S -{A\over6}i\beta$, which only acts on
$\Im s$. This invariance is apparent in eq.~(\ref{arg}).
Secondly, apart from the last term in potential
(\ref{difference}), ${\cal L}_{chiral}$ only depends
on quantities which are invariant under a phase rotation of $m$
or a shift of $\Im s$, like $(m\ov m)$, $\partial_\mu\log(\ov m/m)$,
$s+\ov s$ or $\partial_{\mu}(\Im s)$. The effective theory
${\cal L}_{chiral}$ is then completely insensitive to the choice
of the phase of $\langle m\rangle$: it is the r\^ole of $\Im s$ to cancel
any anomalous dependence on the phase of the gaugino condensate.
The minimum equation (\ref{arg}) ensures, then,
that there exists a vacuum
for all choices of the phase of the gaugino condensate, and that the
physics of the effective theory does not depend on this phase.
In the dual effective lagrangian ${\cal L}_{linear}$,
this information is hidden
in the use of the real vector superfield $V$ for describing
the condensate. As a consequence, all terms in ${\cal L}_{linear}$ are
explicitly invariant under a phase rotation of $m$, and the potential
(\ref{V1}) only depends on $m\ov m$.

It appears then that the PQ symmetry (\ref{Sshift}) is not broken by
the potential for $\Im s$.  This is no surprise, since this symmetry
is actually
used to construct the $U$ dependence of the original 2PI action
(\ref{result}). The same is not true in more general cases,
such as for several condensates which we
consider next, where our duality construction
nevertheless applies equally
well, even though the symmetry (\ref{Sshift}) can
be broken\footnote{A recent
attempt to describe gaugino condensation with a linear multiplet
\cite{GL} only applied to the simplest case of one gaugino condensate.
This corresponds to an effective theory for $S$ with a simple
exponential superpotential. The possibility
of an overall exponential superpotential
compatible with symmetry (\ref{Sshift})
was already pointed out in refs.~\cite{FGKVP,BFQ,FV,DQQ}.
We are also aware that P.~Bin\'etruy, M.K.~Gaillard and T.~Taylor
are making progress along a similar direction.}.

It is instructive to see how the usual conclusions concerning supersymmetry
breaking in this scenario emerge in the linear-multiplet formulation.
As is well known, unlike the case for constant gauge couplings \cite{VY}, a
nonvanishing value of the gaugino condensate necessarily breaks supersymmetry.
We see this in the linear multiplet approach because the gaugino condensate is
represented by the field $m$, which is not the first
component of a superfield. As
a result any nonvanishing value for $m$ must necessarily break supersymmetry.
The same is not so in the chiral representation, where
the condensate is the lowest
component of a supermultiplet. In the chiral case, supersymmetry breaking by
gaugino condensation instead emerges because a nonzero condensate generates a
scalar potential for $s$ which would not be minimized at zero energy.
(This is reminiscent of the discussion in \cite{effsugra,DQQ}\ where
a similar conclusion
was reached in a toy model.)
Notice, however, that a nonvanishing condensate
is not obtained in either theory because of the runaway behaviour of both
potentials, which drives the gaugino condensate to zero.

\section{Several condensates}

The extension of our results to a semi-simple gauge group\footnote{
Abelian factors are not aymptotically-free.} and with one linear multiplet
is simple and interesting.
The microscopic lagrangian is again (\ref{Llagr}),
but with
$$
\hat L = L -2\sum_a c_a \Omega_a,
$$
the index $a$ labelling the simple group factors. The positive constants
$c_a$ specify the normalization of the unified (Wilson) couplings,
$g_a^{-2} = 2c_a\Phi_x$. The equivalent
lagrangian ${\cal L}_V$, as in eq.~(\ref{lagr3}), is
$$
{\cal L}_V = 2\mu^ 2\Dint \Phi(V/\mu^2) + \left({1\over2}\Fint S
\left\{ \sum_a c_a{\Tr}_a (W_a^\alpha W_{a\,\alpha})
+{1\over2}\ov{\cal DD}V \right\}+{\rm h.c.} \right).
$$
And the effective lagrangian (\ref{Lefflin}) becomes
\beq
\label{2leff}
\begin{array}{rcl}
{\cal L}_{eff.} &=& \Dint \left\{2\mu^2\Phi(V/\mu^2)
+\sum_a h_a(U_a\ov U_a)^{1/3} \right\}
\crbig
&&
+\left(\Fint \left\{{1\over2}S(\sum_a c_aU_a + {1\over2}\ov{\cal DD}V)
+{1\over12}\sum_aA_a U_a\log \left( U_a/M^3\right)\right\} +{\rm h.c.}
\right).
\end{array}
\eeq
The $S$-dependent term generates as before the anomaly of the PQ symmetry.
Integration over $S$ imposes the constraint
\beq
\label{2Uis}
\sum_a c_aU_a =  -{1\over2}\ov{\cal DD}V,
\eeq
which generalizes eq.~(\ref{Uis}) and corresponds to
$-{1\over2}\ov{\cal DD}\hat L = \sum_a c_a{\Tr}_a
(W^\alpha_a W_{a\,\alpha})$, which is valid in the microscopic theory.

Before proceeding to analyze the vacuum structure
of this model, recall first the
results for the chiral version of the theory.  The chiral version is
obtained by dualizing the above results, giving
\beq
\label{2Leffdual}
\begin{array}{rcl}
{\cal L}_{chiral} &=&
\Dint\left[ \mu^2 K(S+\ov S) +\sum_a h_a(U_a\ov U_a)^{1/3}
+(S+\ov S)V \right]
\crbig
&& +\left( {1\over12}\sum_a A_a\Fint U_a\log
\left(U_a/M^3\right) +{\rm h.c.} \right),
\end{array}
\eeq
where the vector superfield $V$ is related to the chiral $U_a$ by the
constraint (\ref{2Uis}). The anomaly of the PQ symmetry (\ref{Sshift})
has been transformed into the invariant term $(S+\ov S)V$ by a partial
integration.

The scalar potential for the chiral version of the theory,
(\ref{2Leffdual}), taking (\ref{2Uis}) into account, is
\beq
\label{2V1}
\begin{array}{rcl}
V &=& {\displaystyle
\mu^2 K_{s\ov s} f_s\ov f_s + {1\over9} \sum_a h_a (u_a \ov u_a)^{-2/3}
f_a \ov f_a }
\crbig
&=& {\displaystyle
\sum_a{9\over 4h_a}(u_a\ov u_a)^{2/3}\left|c_a s +
{A_a\over6} \left[1+\log\left({u_a\over M^3}\right) \right]
\right|^2 + {1\over4}\mu^{-2} K_{s\ov s}^{-1}
(\sum_a c_a u_a)(\sum_b c_b \ov u_b)},
\end{array}
\eeq
where $u_a$ and $f_a$ are the lowest and highest components of $U_a$.
This potential is the sum of non-negative terms, and so is minimized when each
vanishes. The condition for the vanishing of each $f_a$ is
\beq
\label{sasphase}
u_a = M^3 e^{-1}  e^{-6c_a s/ A_a}
= M^3 e^{-1} e^{-16\pi^2 s/ C(G_a)},
\eeq
where we recall that the gauge coupling is $g_a^{-2}=2\Re s$, as in eq.
(\ref{couplS}). This establishes the size of each condensate as a function of
the scalar field $s$. The other condition for minimizing the potential is the
vanishing of the auxiliary field $f_s$, which gives:
\beq
\label{allcond}
\sum_a c_a  u_a  = 0,
\eeq
Notice that this sum can be zero without requiring
each of the condensates, $u_a$,
to separately vanish, because $\Im s$ --- which controls
the phases of the $u_a$'s through (\ref{sasphase}) --- can
adjust to permit their cancellation.  But given
our experience with the linear multiplet, for which no
analogue of $\Im s$ exists in
the scalar potential, it would appear to be problematic
to similarly ensure the cancellation of the $u_a$'s. As we
shall see shortly, however, a careful analysis reveals similar conditions
for the linear multiplet.

In order to perform the analysis for the linear multiplet, we
have to use two of the components
of constraint (\ref{2Uis}):
\beq
\label{sumconstr}
\sum_a c_a u_a = - m, \qquad \mbox{and} \qquad \sum_a c_a f_a =
-2 d + i \partial^\mu v_\mu,
\eeq
Unlike the single condensate case, we cannot use these relations to
eliminate all the auxiliary fields $f_a$ as
functions of $d$ and $\partial^\mu v_\mu$. Instead, we can use them
to eliminate $m$ and $d$ in terms of $u_a$ and $R_a\equiv
\Re f_a$ and reintroduce the Lagrange multiplier
$\Im s$ (the axion) to impose the restriction on $I_a\equiv \Im f_a$.
With this information we can write the bosonic part of the lagrangian
(\ref{2leff}) as:
\beq
\label{newlbos}
\begin{array}{rcl}
{\cal L}_{bos.}& = &
{1\over2}\mu^{-2}\Phi_{xx}\left[ \left|
\sum_a c_a u_a\right|^2 + v^\mu v_\mu - (\partial_\mu
c)(\partial^\mu c)\right] +\sum_a K_a(\partial_\mu u_a)(\partial^\mu
\ov u_a)-\Phi_x \sum_a c_a R_a  \crbig
&&  + \sum_a K_a \left(R_a^2+I_a^2\right)-2\sum_a\left(
R_a\Gamma_a+I_a\Lambda_a\right)
+\Im s\left(\sum_a c_a I_a-\partial_\mu v^\mu\right)
\end{array}
\eeq
Where we have  set $K_a\equiv {h_a\over 9}\left(u_a\ov u_a\right)^{-2/3}$
and
we had decomposed the derivatives of the superpotential
$W\equiv \sum_a A_a U_a\log \left( U_a/M^3\right)$ in
their  real and imaginary parts: $W_a\equiv \Gamma_a+i\Lambda_a$.
We can now solve for the auxiliary fields $R_a, I_a,\Im s$
leading to a scalar potential\footnote{Had we solved for the
field $v^\mu$ instead of $\Im s$ we would have recovered
the  potential for  $\Im s$ which
is  the axion field in the chiral version.}:
\beq
\label{newpotl}
\begin{array}{rcl}
V_{linear}&=& - \, {1 \over 2} \mu^{-2} \Phi_{xx} \left|
\sum_a c_a u_a\right|^2+\sum_a K_a^{-1}\left(\Gamma_a+{1\over 2}
\Phi_x c_a\right)^2\crbig
&&
+Q^{-2}\sum_a K_a^{-1}\left[\sum_b K_b^{-1}c_b\left(c_b\Lambda_a-
c_a\Lambda_b\right)\right]^2,
\end{array}
\eeq
with $Q\equiv \left(\sum_d c_d^2K_d^{-1}\right)$, and to the following
derivative couplings for the $v^\mu$ field.
\label{deriv}
\beq
{\cal L}_{der.}=Q^{-1}\left[(\partial_\mu v^\mu)^2
-2\sum_a K_a^{-1} c_a \Lambda_a (\partial_\mu v^\mu)\right]
\eeq
Since the potential is the sum of three positive terms,
 its minimum corresponds
to the points where each of the three terms  vanishes.
This solutions is of course supersymmetric.
Vanishing of the second term fixes the magnitude of $u_a$,
vanishing of the third term implies $\Lambda_a=\xi c_a$ with
$\xi$ arbitrary, this implies:
\beq
\label{ualinear}
u_a=M^3e^{-1} e^{-{c_a\over 2A_a}(\Phi_x-2i\xi)}
\eeq
Finally vanishing of the first term in (\ref{newpotl})
fixes $\Phi_x$ and $\xi$, reproducing the situation of the
chiral case in (\ref{allcond}).

Therefore we have seen that the structure of the scalar potential
is identical in the two dual theories,
in the sense that they both give the same vacuum
with nonvanishing gaugino condensates and
unbroken supersymmetry as  long as
$\sum_a c_a u_a=0$. For the linear multiplet case this conclusion follows quite
naturally, since it is only this particular susy-breaking combination that
corresponds to the $\theta\theta$ component of a superfield. This precisely
corresponds to the situation in the dual chiral version
for which the potential can
develop a minimum  at $\sum_a c_a u_a=0$ without breaking supersymmetry
\cite{krasnikov}.
 This is true even though
in the linear version there is no axion field
developing a potential and therefore a mass, as
it happens  in the chiral case.

Also one can easily see from (\ref{2V1}) that,
as anticipated, the PQ symmetry
(\ref{Sshift}) is {\it not} conserved in the presence of
several condensates. In fact the
first term in (\ref{2V1}) is invariant under (\ref{Sshift})
if we make the shifts
$u_a\rightarrow u_a\exp\{-i (16\pi^2 \alpha/C(G_a))\}$.
However since the shifts are
$G_a$-dependent, the second term in (\ref{2V1})
is not invariant under the shift
(\ref{Sshift}).
This explains why the potential develops a
mass term for the axion field.
 It is remarkable, then, that the theory in this case
nevertheless has a dual version.

It is instructive to ask how this propagating massive degree of
freedom looks in the dual theory. This is particularly so considering that
the usual mechanism for giving mass to an antisymmetric tensor
gauge field --- the Higgs mechanism whereby it and an ordinary
spin-one gauge field combine into a massive spin-one particle --- is
not available in the present case. Inspection of the previous
formulae shows that this degree
of freedom is described by the vector field $v^\mu$,
in the linear-multiplet version, and that it
describes a massive spinless particle in a somewhat unusual way.

These conclusions follow from the lagrangian for $v^\mu$,
which is
\beq
\label{lagrangvmu}
 {\cal L}_{v^\mu}={1\over2}\mu^{-2}\Phi_{xx} v^\mu v_\mu+
Q^{-1}\left[(\partial_\mu v^\mu)^2
-2\sum_a K_a^{-1} c_a \Lambda_a (\partial_\mu v^\mu)\right],
\eeq
with the coefficients of each term now taken to be
the value of the corresponding function in the vacuum. Clearly, this
lagrangian describes a propagating massive scalar degree of
freedom given by the longitudinal part of $v^\mu$ (the
transverse spin one components  do not propagate) corresponding
to the axion. The only component of this field whose time derivatives appear
in ${\cal L}_{v^\mu}$ is $v^0$, so the other three components
can be considered to be auxiliary fields.
The  propagator for $v^\mu$ which follows from ${\cal L}_{v^\mu}$ can
be computed, and is (taking, for simplicity, ${1\over2}\mu^{-2}\Phi_{xx}=-1$),
$\delta_{\mu\nu}-{k_\mu k_\nu\over Q+k^2}$. A similar propagator for
the massive axion has also recently been used by R. Kallosh {\it et al}
\cite{KLLS}\ in their discussion of the axion mass ($m^2=Q$).
We have, therefore, a lagrangian
description of a massive axion in terms of a vector field or,
equivalently, a massive 3-index antisymmetric tensor field.

Supersymmetrically, the vector superfield $V$ which (off-shell) had
eight bosonic degrees of freedom,  has only two bosonic
degrees of freedom on-shell, corresponding to the massive
scalar fields $c$ and the longitudinal component of
$v^\mu$. The three transverse components of $v^\mu$
as well as $m, \ov m$ and $d$
do not propagate. We are led to a formulation for which,
after gaugino condensation,
the original $b_{\mu \nu}$ field of the
linear multiplet is projected out of the spectrum in favour
of a massive scalar field inside $v^\mu$. This solves the
puzzle of the axion mass, and gives a novel mechanism for $b_{\mu\nu}$
to acquire a mass.

\section{Conclusions}

We have shown that, contrary to previous belief, it is possible to
formulate gaugino condensation directly in the linear multiplet formalism.
This permits the analysis to be performed using the multiplet in which
string theory presents the dilaton. It is our hope that the existence of this
alternative point of view may lead to new insights into the physics of
dynamical supersymmetry breaking in string-like theories.

A possible advantage of using the linear multiplet in
string theory is that it is the function $\Phi(L)$ which is directly obtained
from perturbative string calculations. Working directly with $\Phi(L)$
obviates the necessity for eliminating $V$ as a function of $S+\ov S$, as is
required to determine the lagrangian in its chiral form. As may be seen from
equation (\ref{VLeom}),  this can usually {\it not} be done exactly.

We show the result which we obtain to be dual to the standard chiral approach,
even though the PQ symmetry on which this duality is based in the microscopic
theory is broken, in general, in the low energy lagrangian. We find in so doing
a
novel mechanism for giving a mass to an antisymmetric tensor gauge field.
As a result, we  give the
first good evidence that chiral-linear duality is an exact symmetry of string
theories, which survives strong coupling effects.

\section*{Acknowledgements}

We wish to thank L.~Alvarez-Gaum\'e  for useful discussions.
One of us (C.B.) also wishes to thank V.
Kaplunovsky for useful conversations.



\begin{thebibliography}{9}

\bibitem{AKMRV}
For a review, see:
D. Amati, K. Konishi, Y. Meurice, G. C. Rossi and G. Veneziano,
{\it Phys. Rep.} {\bf 162} (1988) 169

\bibitem{VY}
G. Veneziano and S. Yankielowicz, \PL{B113}{82}{231}

\bibitem{others}
T. R. Taylor, G. Veneziano and S. Yankielowicz, \NP{B218}{83}{493};
\\
K. Konishi and G. Veneziano, \PL{B187}{87}{106}

\bibitem{DIN}
J.-P. Derendinger, L. E. Ib\'a\~nez and H. P. Nilles, \PL{B155}{85}{65}

\bibitem{DRSW}
M. Dine, R. Rohm, N. Seiberg and E. Witten, \PL{B156}{85}{55}

\bibitem{condens2}
N. P. Nilles, \PL{B115}{82}{193}; \NP{B217}{83}{366};\\
S. Ferrara, L. Girardello and H. P. Nilles, \PL{B125}{83}{457}

\bibitem{KP}
C.~Kounnas and M.~Porrati, \PL{B191}{87}{91}

\bibitem{linear}
S. Ferrara, J. Wess and B. Zumino, \PL{B51}{74}{239}; \\
W. Siegel, \PL{B85}{79}{333}

\bibitem{22}
L. Dixon, V. S. Kaplunovsky and J. Louis, \NP{B355}{91}{649};\\
I. Antoniadis, K. S. Narain and T. R. Taylor, \PL{B267}{91}{37};\\
I. Antoniadis, E. Gava and K. S. Narain, \PL{B283}{92}{209};
\NP{B383}{92}{93};\\
P. Mayr and S. Stieberger, \NP{B407}{93}{725}; \NP{B412}{94}{502}; \\
D.~Bailin, A.~Love, W.A.~Sabra and S.~Thomas, \PL{B320}{94}{21};
{\it Mod. Phys. Lett.} {\bf A9} (1994) 67;\\
E.~Kiritsis and C.~Kounnas, preprint CERN-TH.7472/94,
hep-th/9501020

\bibitem{effsugra}
J.-P. Derendinger, S. Ferrara, C. Kounnas and F. Zwirner,
\NP{B372}{92}{145}; \PL{B271}{91}{307}; \\
P. Bin\'etruy, G. Girardi and R. Grimm, \PL{B265}{91}{111};\\
P. Adamietz, P. Bin\'etruy, G. Girardi and R. Grimm,
\NP{B401}{93}{257};\\
G.~Lopes Cardoso and B.~Ovrut, \NP{B369}{92}{351};
\NP{B392}{93}{315}


\bibitem{DQQ}
J.-P. Derendinger, F. Quevedo and M. Quir\'os, \NP{B428}{94}{282}

\bibitem{BDQQ}
C. P. Burgess, J.-P. Derendinger, F. Quevedo and M. Quir\'os, to appear
(January 1995)

\bibitem{CJT}
J. M. Cornwall, R. Jackiw and E. Tomboulis, \PR{D10}{74}{2428};\\
T. Banks and S. Raby, \PR{D14}{76}{2182};\\
See also: M. Peskin, in `Recent Developments in Quantum Field
Theory and Statistical Mechanics', Les Houches 1982, ed. by J.-B. Zuber
and R. Stora (North-Holland, Amsterdam, 1984)

\bibitem{T}
T. R. Taylor, \PL{B164}{85}{43}

\bibitem{G}
S. J. Gates Jr., \NP{B184}{81}{381}

\bibitem{GW}
M. T. Grisaru and P. C. West, \NP{B254}{85}{249}

\bibitem{GL}
I.~Gaida and D. L\"ust, preprint HUB-IEP-94/33, hep-th/9412079

\bibitem{FGKVP}
S. Ferrara, L. Girardello, S. Kugo and A. Van Proeyen, \NP{B223}{83}{191}

\bibitem{BFQ}
C.~Burgess, A.~Font and F.~Quevedo, \NP{B272}{86}{661}

\bibitem{FV}
S.~Ferrara and M.~Villasante, \PL{B186}{86}{85}

\bibitem{krasnikov}
N.V.~Krasnikov, \PL{B193}{87}{37}

\bibitem{KLLS}
R. Kallosh, A. Linde, D. Linde and L. Susskind,
preprint SU-ITP-95-2, hep-th/9502069.
\end{thebibliography}
\end{document}